\documentclass[floatfix,aps,prl,twocolumn,twoside,preprintnumbers,amsmath,amssymb,longbibliography]{revtex4-1}

\usepackage{graphicx}
\usepackage{dcolumn}
\usepackage{bm}
\usepackage{bbm}
\usepackage{mathrsfs}
\usepackage{pifont}
\usepackage{tabularx}
\usepackage[utf8]{inputenc}

\usepackage{amsmath,amsfonts,amssymb}
\usepackage[usenames]{color}
\usepackage[bookmarks, colorlinks=true, plainpages = false, citecolor = blue, linkcolor = blue, urlcolor = blue, filecolor = blue]{hyperref}

\usepackage{xcolor}
\usepackage{enumerate}

\usepackage{tikz}

\newcommand{\gd}{\ensuremath{\dot\gamma}}
\newcommand{\tg}{\ensuremath{T_g}}
\newcommand{\ie}{{ \it i.e.,~}}

\newcommand{\etal}{{ \it et al.~}}

\newcommand{\sxy}{\ensuremath{\sigma_{xy}}}

\newcommand{\beq}{\begin{equation}}
\newcommand{\eeq}{\end{equation}}

\makeatletter
\renewcommand*{\fnum@figure}{{\normalfont\bfseries Fig.~\thefigure}}
\renewcommand*{\@caption@fignum@sep}{\textbf{\ }}
\makeatother
\makeatletter
\renewcommand{\thetable}{\arabic{table}}
\renewcommand*{\fnum@table}{{\normalfont\bfseries Table~\thetable}}
\makeatother

\begin{document}


\title{ Active microrheology of a bulk metallic glass}

\author{Ji Woong Yu$^1$}
\author{S. H. E. Rahbari$^2$} \email{habib.rahbari@gmail.com}
\author{Takeshi Kawasaki$^3$}
\author{Hyunggyu Park$^2$}
\author{Won Bo Lee$^1$}\email{wblee@snu.ac.kr}

\affiliation{$^1$ School of Chemical and Biological Engineering,
  Institute of Chemical Processes, Seoul National University, Seoul
  08826, Korea}

\affiliation{$^2$ School of Physics, Korea Institute for Advanced
  Study, Seoul 02455, Korea}

\affiliation{$^3$ Department of Physics, Nagoya University, Nagoya
  464-8602, Japan }

\begin{abstract}
	
	The glass transition remains unclarified in  condensed matter physics. Investigating the mechanical properties of glass is  challenging because any global deformation that may result in shear rejuvenation requires an astronomical relaxation time. Moreover, it is well known that a glass is heterogeneous and a global perturbation cannot explore local mechanical/transport properties. However, an investigation based on a local probe, \ie microrheology, may overcome these problems. Here, we establish active microrheology of a bulk metallic glass:
	a probe particle driven into host medium glass. This is a technique amenable for experimental investigations. We show that upon cooling the microscopic friction exhibits a second-order phase transition; this sheds light on the origin of friction in heterogeneous  materials. Further, we  provide distinct evidence to demonstrate that a strong relationship exists between the microscopic dynamics of the probe particle and the macroscopic properties of the host medium glass. These findings establish active microrheology as a promising technique for investigating  the local properties of bulk metallic glass. 

\end{abstract}

\maketitle

\noindent{\large\bf Introduction\par}
When a glass-forming liquid is rapidly cooled below its glass transition temperature, it falls out of equilibrium, and its viscosity, $\eta$, increases by more than fifteen orders of magnitude~\cite{gotze_1992, angell_1995}.  Notably, the structural relaxation time of a glass, $\tau_\alpha \propto \eta$, exceeds any other time scale in the realm of condensed matter physics and is comparable to cosmological time scales~\cite{sethna_1991}. The origin of this slowing down remains the primary challenge in the physics of glass transition: whether the long time scale originates from collective behavior   (a thermodynamic description)~\cite{kardar_2007}, similar to critical phenomena, or the slowing down occurs at the microscopic level (a kinetic origin). A  widely accepted picture about slow progress of a supercooled liquid towards equilibrium is based on a trap-escape behavior: a particle is transiently trapped in a cage created by its neighbors; over a  short time, $\tau_\beta$, the particle rattles inside the cage, escaping only after a long time, $\tau_\alpha$. This time scale diverges at the glass transition temperature, where the cages become permanent. This interpretation is well supported by the characteristic two-step relaxation of a supercooled liquid, as observed in neutron scattering experiments and simulations~\cite{binder_2011}. Several competing theories aimed at explaining the origin of the slowing down of a glass. Among those theories, mode-coupling theory (MCT) has successfully described power-law $\beta$-relaxation, stretched exponentials of $\alpha$-relaxation, and the time-temperature superposition principle~\cite{janssen_2018}. An intriguing discovery in experiments~\cite{russell_2000, ediger_2000} and simulations~\cite{hurley_1995, yamamoto_1998} has motivated a thermodynamic description of the glass transition: localized particles, unable to diffuse away, form mesoscopic clusters whose typical size increases upon cooling~\cite{tanaka_2010}. These mesoscopic structures, known as  dynamic heterogeneities (DHs), whose origin is yet another controversy (static versus dynamic origin), suggest that a supercooled liquid can be treated as a complex fluid~\cite{furukawa_2009}. This interpretation is in contrast to the original MCT which was an extension of simple the liquid theory; although recent developments of MCT reconciled with DHs~\cite{biroli_2006}. In addition to these scenarios, amorphous order~\cite{albert_2016}, random-first-order transition~~\cite{kirkpatrick_2015}, and frustration~\cite{tarjus_2005, shintani_2006} have also been discussed extensively as possible explanations for the origin of the glass transition.\\

  Despite intense research over the  last decades on the underlying origins of the glass transition, a consensus has not yet been established. Furthermore, an understanding of the mechanical properties of glasses, which are investigated by shear  deformation, is also elusive. The reason for this lack of consensus is four-fold: ({\bf 1}) any global perturbation requires a dramatically long time scale for relaxation, ({\bf 2}) shear forces may rejuvenate the glass, relocating the glass to shallower energy minima~\cite{parmar_2019, lacks_2004} and rendering the glass more prone to fluctuations, ({\bf 3}) a realistic deformation rate for an  atomistic glass is a technical impossibility in computer simulations~\cite{berthier_2002, voigtmann_2011}, and ({\bf 4}) due to DHs a glass is heterogeneous and a global deformation only explores an average mechanical response. Here, we overcome these problems by investigating the mechanical properties of a glass at the microscopic level. Using this technique, known as active microrheology, a singled-out probe particle is pulled through a dense host liquid~\cite{zia_2018}. Investigation of the  probe particle dynamics reveals non-equilibrium properties of the host medium, such as viscosity. \\

\begin{figure*}
	\includegraphics[width=7.0in]{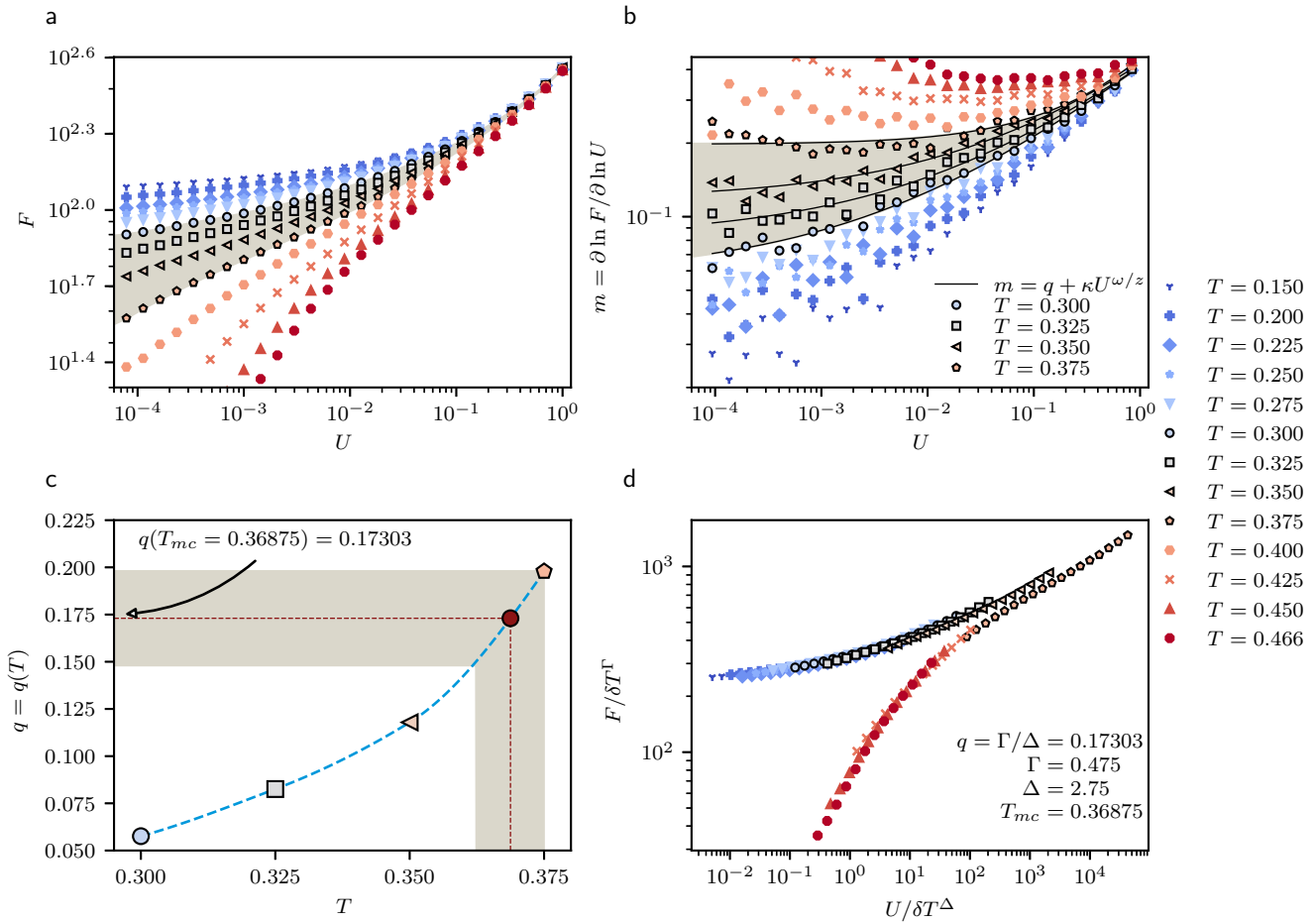}
	\caption{{\bf Velocity-force curves.} ({\bf a}) Friction, $F$, versus the probe velocity, $U$, for various temperatures across the  glass transition temperature. Each point is an ensemble average over $500$ independent realizations. ({\bf b}) Successive slope of the velocity-force curves $m = \partial \ln F / \partial \ln U$ versus the probe velocity. A negative-positive cross over of the curvature occurs in the range of $T = 0.375$ to $0.300$, marking the critical transition region. Equation~\eqref{eq:succ_slope} is fitted to each curve to compute the corresponding asymptotic exponent $q$ in the critical region (solid lines). ({\bf c}) The asymptotic exponent $q$ varies monotonically with temperature $T$ across the transition region. The data are interpolated by splines, which enables us to obtain the asymptotic exponent for any temperature within the transition region. The shaded region  shows the range over which the data can be collapsed. ({\bf d}) An excellent collapse is obtained  for $T_{mc} = 0.36875$, with an asymptotic exponent of $q = \Gamma / \Delta = 0.17303$ in which $\Gamma =0.475$ and $\Delta = 2.75$.  
		\label{fig:force_velo_scaling}}
\end{figure*}

\noindent{\bf Microrheology.} We perform active microrheology of a probe particle with a constant velocity $U$. Density map snapshots of the system are given in Supplementary Fig. 1. The size of the probe particle mimics that of a typical length scale of DHs. We obtain the response of the host medium by measuring the force acting on the particle, \ie the friction, $F = -{\bf F}\cdot {\bf \hat{n}}$, where ${\bf \hat{n}} = {\bf U} / |{\bf U} |$. The host medium is modeled by a  Kob-Andersen mixture in $3$D, whose thermodynamic glass transition temperature, the Kauzmann temperature,  is $T_K = 0.3$ (more information regarding the model is given in Methods section). Before performing microrheology simulations, we used a state-of-the-art method to deeply anneal the host medium  with cyclic shearing~\cite{leishangthem_2017}. This step minimizes the occurrence of plastic events, rendering the system less prone to fluctuations. In Fig.~\ref{fig:force_velo_scaling}a, the magnitude of the frictional force acting on the probe particle, $F$, is depicted versus the probe velocity $U$. Each color corresponds to a different temperature $T$. A complicated nonlinear velocity-force relationship is found. The velocity-force curves can be separated into two distinct classes. In the first class, at high $T$,  the friction  $F$ vanishes as $U \to 0$.  However, in the second regime, at low $T$, the force $F$ is finite at $U \to 0$. To our best knowledge, this is the first report of such behavior for a bulk metallic glass, although a similar threshold has been found for soft colloids \cite{gazuz_2009, hastings_2003}. This threshold force is akin to the yield stress of a paste~\cite{bonn_2017}. The emergence of a non-zero threshold force $F$ is reminiscent of a phase transition and can be interpreted as an arrest transition: the probe particle will not delocalize if the probe is driven by a force smaller than the threshold force. Interestingly, this crossover from delocalization to an arrest regime occurs at low $T$, where the glass transition occurs. Therefore, 
there may be an intimate relationship between the microscopic dynamics of the probe particle and the macroscopic properties of the host medium. We will demonstrate existence of such relationship in the remainder of this paper. The crossover region from the arrest regime to delocalization, which is marked by the shaded region in Fig.~\ref{fig:force_velo_scaling}a, is remarkably wide. We will demonstrate that this wide region is a beneficial advantage of microrheology.   \\



\noindent{\bf Scaling ansatz.} To fully capture the nonlinear behavior of the velocity-force curves, we seek a scaling ansatz reminiscent of that for magnetization in an Ising model. In this approach, the friction, $F$, depends only  on a scaling variable $x = U /  |T - T_{mc}| ^ \Delta$ via 
\begin{equation}
F(U, T) = |T - T_{mc} | ^\Gamma \mathbb{F} \left( \frac{U} { |T - T_{mc}| ^ \Delta} \right),
\label{eq:F_U_scaling}
\end{equation}
in which $\delta T = |T - T_{mc} |$ is the distance from a critical delocalization temperature $T_{mc}$, $\mathbb{F}(\dots)$ is a scaling function, $\Gamma$ is a parameter-dependent exponent, and $\Delta$ is a universal gap exponent. At $T = T_{mc}$, scale invariance requires $F \propto U ^ q$, in which $q$ is a critical exponent, and  Eq.~\eqref{eq:F_U_scaling} confers $q = \Gamma / \Delta$ (a formal derivation is given in Supplementary Note~1). We previously developed a systematic framework to extract the critical density and critical exponent for the macrorheology of jamming~\cite{rahbari_2018}. Here, we  generalize this framework for microrheology to extract $T_{mc}$ and the critical exponent $q$. This technique is based on the successive slope of the velocity-force curves, defined as $m = {\partial \ln F }/{\partial \ln U}$: at $T= T_{mc}$, the curvature of the $m-U$ curves crosses over from negative to positive. This fact can be used as a criterion to find critical region. As shown in Fig.~\ref{fig:force_velo_scaling}b, we computed the successive slope of the velocity-force curves, $m$, for various temperatures. The slope $m$ can be analytically calculated from Eq.~\eqref{eq:F_U_scaling}, and at $T = T_{mc}$, the slope becomes $m=q$. From \mbox{Fig. ~\ref{eq:F_U_scaling}b}, we can see that the curvature of $m-\gd$ curves crosses over from negative to positive in the range of $T = 0.300$ to $0.375$. Clearly, at $T = 0.375$ (highest bold symbols), the data curve upward in the asymptotic limit of $U \to 0$;  similarly, at $T = 0.300$ (lowest bold symbols), the data curve downward for the same limit. Therefore, the critical temperature must be in the range $0.300 < T < 0.375$. We also observe a strong correction-to-scaling for $U > 10^{-2}$, indicating that at $T = T_{mc}$, a correction-to-scaling of the form $F = U ^ q (\kappa_1 + \kappa_2 U ^{\omega/z})$ is required, in which $\omega$ is the leading correction-to-scaling exponent, and $z$ is the dynamic exponent. Therefore, at $T =T_{mc}$, the successive slope becomes
\begin{equation}
m = q + \kappa U^{\omega/z}.
\label{eq:succ_slope}
\end{equation} 

Because the critical temperature $T_{mc}$ is unknown,  we fit Eq.~\eqref{eq:succ_slope} to the data in the critical region, $0.300 < T < 0.375$. The resulting lines are depicted by solid curves in Fig.~\ref{fig:force_velo_scaling}b, and we display the values of the fit parameters in Table~\ref{tab:fits}.\\

\begin{table}
	\[\begin{array}{cccc}
	\hline \hline
	\multicolumn{1}{c}{\text{Temperature}} & \multicolumn{1}{c}{\text{$q$}} & \multicolumn{1}{c}{\text{$\kappa$}} & \multicolumn{1}{c}{\text{$\omega/z$}}\\
	\hline\\[-2mm]
	0.300 & 0.0575 & 0.35 & 0.355 \\[1mm]
	0.325 & 0.0825 & 0.329 & 0.360 \\[1mm]
	0.350 & 0.117 & 0.303 & 0.377 \\[1mm]
	0.375 & 0.197 & 0.245 & 0.621 \\
	\hline\hline
	\end{array}\]
	\caption{ Fitting parameters for
		Eq.~\eqref{eq:succ_slope} displayed by  the solid lines in  Fig.~\ref{eq:F_U_scaling}b.
		\label{tab:fits}}
\end{table}


%

The magnitude of the pre-factor $\kappa$ in Eq.~\eqref{eq:succ_slope} is crucial in this analysis, because it determines the reliability of the fit. For  high-quality data, one must obtain $\kappa \simeq \mathcal{O}(1)$. From Table~\ref{tab:fits}, one can see that $\kappa$ is on the order of unity and does not change dramatically  with temperature. This behavior demonstrates the quality of our data due to the careful preparation of a well-annealed medium, as well as the rarity of shear rejuvenation in microrheology. The asymptotic exponent $q$ changes rapidly with $T$, as illustrated in Fig.~\ref{fig:force_velo_scaling}c. We interpolate the $q-T$ data with splines, which enables us to obtain the corresponding $q$ for any $T$ in the critical region. If $F$ scales according to Eq.~\eqref{eq:F_U_scaling}, then rescaling of the data with ${F}/{\delta T^\Gamma}$ versus ${U} / {\delta T^\Delta}$ must lead to a collapse of the data into a master curve. This rescaling of the flow curves is of canonical interest for both fundamental and technological applications for predicting the flow properties of materials~\cite{dinkgreve_2018, voigtmann_2011}. The ratio of these exponents $q = \Gamma / \Delta$ is obtained through interpolation, and we vary the exponent $\Gamma$ at a given $T$ to achieve a collapse. Remarkably, we only obtain a  collapse of the data within a narrow range of  $T_{mc} = 0.36875 \pm 0.0062$  ( the shaded area in Fig.~\ref{fig:force_velo_scaling}c), with the optimal collapse at the center of the interval. The error bars account for a mere $1.5\%$ uncertainty. The narrowness of the interval over which a collapse can be obtained  is very similar to the results for conventional critical phenomena. In contrast,  flow curves of  athermal system undergoing a jamming transition exhibit an uncertainty of approximately $5\%$~\cite{rahbari_2018}, leading to the long-standing controversy over the exponents involved in jamming~\cite{vagberg_2016}. We conclude that the critical delocalization temperature is $T_{mc} =  0.36875 \pm 0.0062$, which is approximately $23\%$ higher than the thermodynamic glass transition temperature, at $T_K = 0.3$. The delocalization temperature $\tg$ is larger than the thermodynamic glass transition temperature $T_K$, which is consistent with the findings of Habdas\etal\cite{habdas_2004}. The critical exponent associated with this temperature is $q = 0.17303 \pm 0.0248$, and the crossover exponents are $\Gamma =0.475$ and $\Delta = 2.75$. Using these exponents, an excellent collapse is obtained, as shown in Fig.~\ref{fig:force_velo_scaling}d, confirming 
a phase transition for the probe particle. Although we obtain an equilibrium regime at high T, \ie $F \propto U^1$, the velocity-force dependence is otherwise highly nonlinear. Therefore, this collapse is primarily over non-equilibrium region. This behavior is remarkable because critical scaling collapses generally occur in at equilibrium. We also observe that the velocities over which a linear response can be observed decrease as $T_K$ is approached, which is
consistent with  the report of Williams and Evans~\cite{williams_2006}.  \\

\begin{figure*}[!ht]
        \includegraphics[width=7.0in]{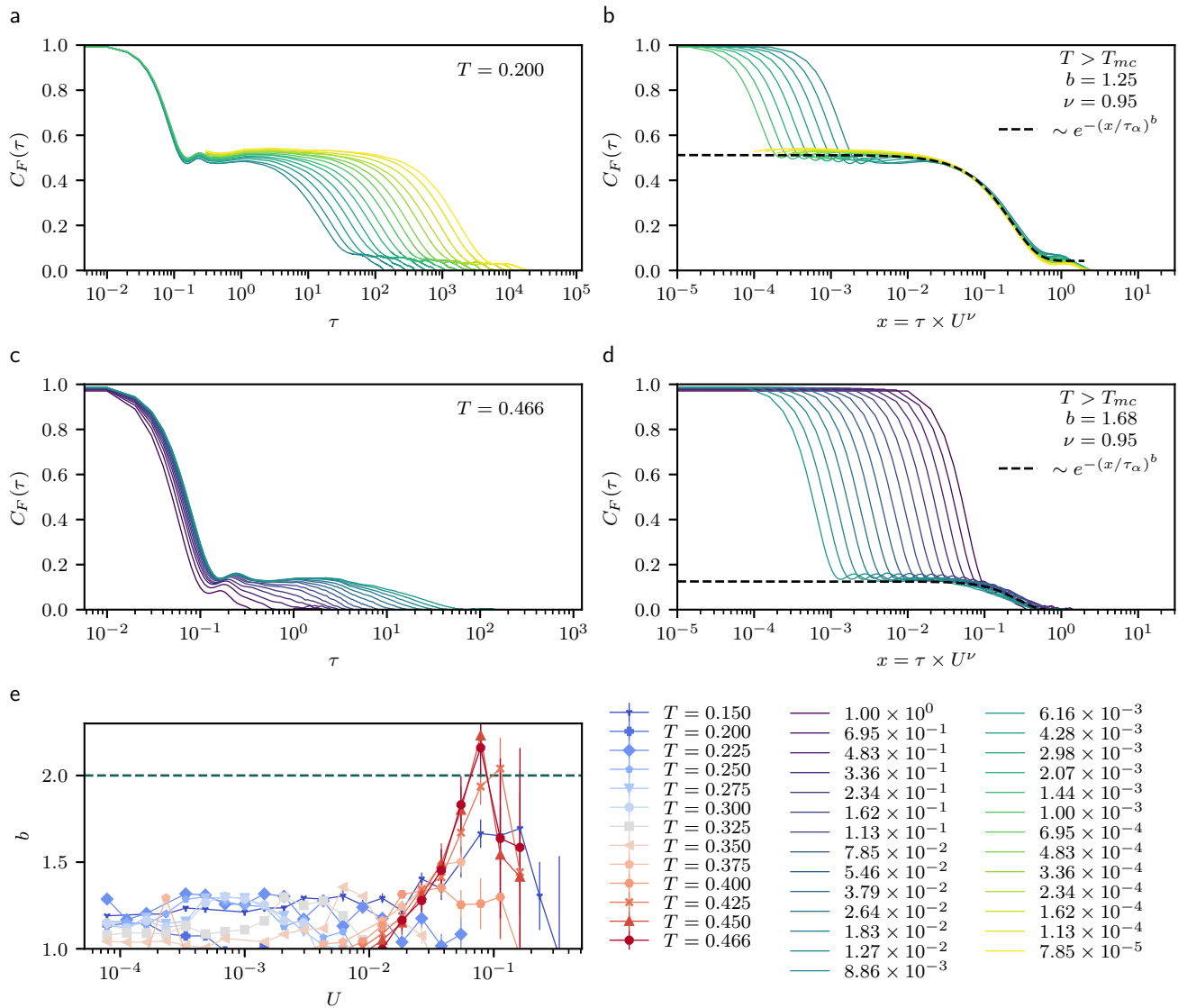}
	\caption{{\bf Two-step relaxation of friction.}
		The auto-correlation function $C_F(\tau) = \langle F(0)F(\tau) \rangle$ of the time
		series of the force acting on the probe particle $F$ versus the time window $\tau$ for $T = 0.200$ and $0.466$ is depicted in panel {\bf a} and {\bf c}, respectively. Each curve represents an average of $500$ realizations. Each curve has an initial relaxation, which corresponds to ballistic transport at very small time scales. This fast relaxation is followed by a plateau, which is akin to a $\beta$-process (rattling in cages). Each plateau starts with a damped 	oscillation, known as  quenched echoes \cite{kob_1997, nagel_1983}, which corresponds to the oscillatory nature of the initial annealing of the system or phonon propagation~\cite{shiba_2016}. The plateau terminates with a long relaxation process akin to an $\alpha$-process (escape).  We plot $C_F(\tau)$ versus $x = \tau \times U ^\nu$, in which $\nu = -(q - 1) = 0.95\pm 0.1$ is the force thinning exponent in panel {\bf b} and {\bf d}. The data collapse onto a compressed exponential function of the form Eq.~\eqref{eq:stretched} (dashed lines in panel {\bf b} and {\bf d}). The collapse is akin to the time-shear superposition principle in spin glass and MCT~\cite{gotze_1992, berthier_2000}. The compressed (faster) exponential is a manifestation of the super-diffusive motion of the probe particle, corresponding to super-diffusive relaxation. In panel {\bf e}, the compressed exponent $b$ is depicted versus the probe velocity for various values of $T$.
		\label{fig:force_acf}}
\end{figure*}

\noindent{\bf Relaxations.} We now focus on the properties of the time series of the force acting on the probe particle $F = \{F(t)\}$, and we investigate whether the structural relaxation process can be traced via a friction time series. In Supplementary Fig.~1a,~b, we display the friction time series at $T = 0.150$ and $0.466$, respectively, for a probe velocity of $U=0.0379269$.  As shown in Supplementary Fig.~1b, in the supercooled state, the friction changes rapidly with time; in contrast, in Supplementary Fig.~1a, corresponding to the glass phase, the friction exhibits a strong correlation. The time interval over which the friction is correlated is marked by an arrow as a guide to the eye. This apparent correlation may be related to the formation of cages. To investigate this possibility, in Fig.~\ref{fig:force_acf}a,~c,  we plot the auto-correlation function for friction, $C_F(\tau)=\langle F(\tau)F(0) \rangle$, for $T = 0.200$, and $0.466$, respectively. $\langle \cdots \rangle$ presents an ensemble average over $500$ independent realizations (in Supplementary Fig.~3 we show a similar plot for $T=0.350$). Each curve begins with an initial relaxation, which corresponds to the ballistic motion of particles at very small time scales. One can see that all different curves with different $U$ values are superimposed for short time scale, indicating the fact that the short-term dynamics is unaffected by the probe motion. This behavior is analogous to that of sheared liquids at very small time scales, which is similarly unaffected by shear~\cite{varnik_2006}. The curves level off with a dip followed by a damped oscillation and eventually  a plateau. The dip and the damped oscillation correspond to the oscillatory nature of the initial system preparation  due to shear annealing~\cite{das_2018}. The damped oscillation of the correlation function is known as quenched echoes \cite{nagel_1983}. Kob and Barrat reported such oscillations for Lennard-Jones (LJ) glasses annealed with oscillatory protocols~\cite{kob_1997}. However, Shiba\etal showed similar oscillations originated from propagation of acoustic phonons~\cite{shiba_2016}. These authors ruled out the possibility of this dip being related to the boson peak. After the damped oscillations, a plateau with a length of several orders of magnitude appears. This trend is akin to the $\beta$-relaxation process~\cite{binder_2011}. After this rattling period, a long relaxation process terminates the curves. This final long relaxation is akin to an $\alpha$-relaxation process, corresponding to the escape of a particle from a cage. The behavior of the system both below and above the glass transition temperature is consistent with  that in shear-driven structural glasses~\cite{berthier_2000, varnik_2006}. Therefore, the relaxation of friction is described by a two-step relaxation process. This reveals a strong relationship between dynamics of the probe particle and the two-step structural relaxation of the host medium glass. In the remaining of the paper, we will provide more evidence for this intimate relationship. \\





\noindent{\bf Time-force superposition.} In Fig.~\ref{fig:force_acf}b,~d, $C_F(\tau)$ is plotted versus $x = \tau \times U^\nu$, in which $U$ is the velocity of the probe particle and $\nu = -(q - 1) = 0.95\pm 0.1$ is the force thinning exponent ($q$ is the critical exponent of the scaling ansatz in Eq.~\eqref{fig:force_velo_scaling}); here, an excellent collapse of the data into a master curve is obtained. The collapse of the correlation function is a time-force superposition, which is reminiscent of the "time-shear superposition principle" predicted by both MCT~\cite{gotze_1992} and spin glass theory~\cite{berthier_2000}.
The force thinning exponent $\nu$ is consistent, within the error bars, with the shear thinning exponent reported by Furukawa \etal  $\nu = 0.8$ \cite{furukawa_2009}, very close to the value reported by Berthier-Barrat,  $\nu = 2/3$~\cite{berthier_2002},  and Varnik $\nu =1$~\cite{varnik_2006}. Accordingly, our force thinning exponent, $\nu$, and the critical exponent, $q=1-\nu$, within the error bars, are consistent with findings reported in the literature for macrorheology. \\

\noindent{\bf Compressed exponential.} The master function (dashed lines in Fig.~\ref{fig:force_acf}b,~d) is given by a compressed exponential function of the following form (dashed lines in Fig.~\ref{fig:force_acf}b,~d):
\begin{equation}
C_F(\tau) \simeq e^{ -\left(x / \sigma\right) ^ {b\left(T, U\right)} }.
\label{eq:stretched}
\end{equation}
in which $b(T, U)\ge 1$ is a temperature- and velocity-dependent (compressed) exponent. In Fig.~\ref{fig:force_acf}e, the compressed exponent is plotted versus the probe velocity, $U$, for various temperatures. The exponent is generally larger than unity and reaches a ballistic value of $2$ in the supercooled regime for large $U$ values. This result is in contrast with the stretched-exponential form $b<1$ predicted by the MCT. The faster-than-exponential relaxation $b > 1$ stems from the super-diffusive motion of the probe particle $\langle \delta r_\text{probe}^2 \rangle \propto  t^2 $. This result sheds light on a recent open debate on the origin of the compressed exponential relaxation in  colloidal systems and glass formers~\cite{gnan_2019, angelini_2013}. Another important characteristic of a glass is aging~\cite{kob_1997}: the longer an observer waits, the slower the glass becomes. This characteristic is simply due to particles trapped in the cages.  However, shear rejuvenates a glass and stops aging~\cite{angelini_2011, varnik_2006}: cages are continually crushed by shear rejuvenation. Although glass is at non-equilibrium,  a sheared glass reaches a stationary state similar to equilibrium, and the correlation functions only depend on the time difference. We examined aging in active microrheology with a two-point correlation function, $C_F(\tau, \tau_w)$, in which $\tau_w$ is a waiting time. In agreement with findings for sheared glasses, we found that $C_F(\tau, \tau_w) $ is time-invariant and does not exhibit any aging effects. However, we do not suggest that this behavior is related to shear rejuvenation, as our probe is local. This behavior is expected to arise because the activated probe particle continually breaks the cages and does not get trapped in any cage. \\

\begin{figure}[!ht] 
	\includegraphics[width=3.4in]{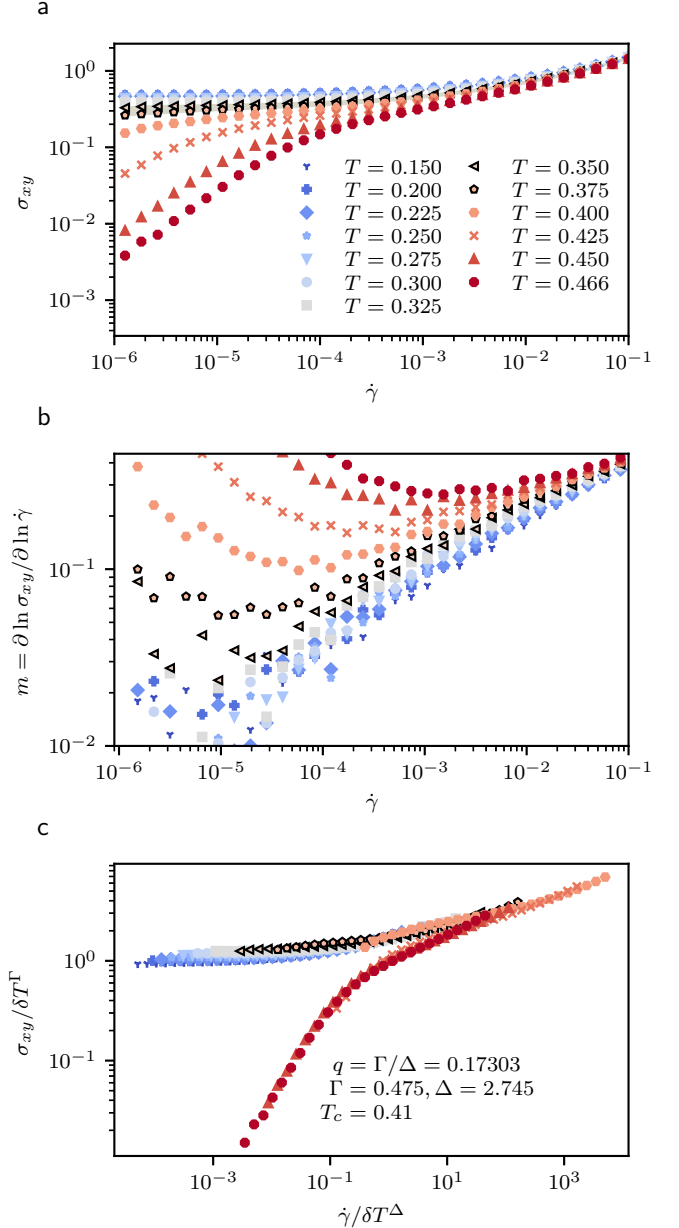}
	\caption{{\bf Flow curves.} ({\bf a}) Shear stress, $\sxy$, versus the shear rate, $\gd$, for different temperatures. In shaded region, a crossover from fluid to solid can be seen in a very narrow range. As a result, the critical region is very small. ({\bf b}) Successive slope of the flow curves, $m$, versus the shear rate $\gd$. In contrast to the successive slope observed for active microrheology, we observe large fluctuations for $U\to 0$. These large  fluctuations prevent a systematic scaling analysis to extract the critical exponents. ({\bf c}) Flow curves collapsed with the exponents obtained from active microrheology at a critical temperature $T_c = 0.41$. The collapse has a lower quality than that for microrheology.
	\label{fig:macrorheo}}
\end{figure}

\noindent{\bf Macrorheology.} A conventional way of exploring the mechanical properties of supercooled liquids is macrorheology. We now compare the results obtained from active microrheology and macrorheology. We performed simulations of a simple shear applied to the host medium by employing the Lees-Edwards boundary conditions. The resulting flow curves are shown in Fig.~\ref{fig:macrorheo}a. For most of the flow curves at low temperature, the shear stress saturates to a shear-rate-independent regime at low $\gd$ . This saturation signals the emergence of yield-stress and a transition from fluid to solid states. Moreover, the crossover from fluid to solid regimes occurs at an extremely narrow range between $T= 0.400$ (squares) and $0.375$ (diamonds). This region is marked by shaded area, whose width can be compared to a similar region in Fig.~\ref{fig:force_velo_scaling}a. The comparison reveals a striking fact:  macrorheology provides a much narrower critical range compared with microrheology. We will see that the narrowness of the transition region makes a systematic scaling analysis nearly impossible for macrorheology. To this end, in Fig.~\ref{fig:macrorheo}b, we plot the successive slope of the flow curves. In contrast to that in microrheology, the successive slope is very noisy, with large fluctuations for $U\to 0$; these fluctuations are characteristic of shear rejuvenation and slow dynamics at low $T$. Similar to the microrheology, the critical region is defined as the range at which the curvature of $m-\gd$ data cross over from positive to negative  (black). This happens in the range between $T= 0.400$ and $0.375$. A systematic analysis, similar to that for microrheology, is not possible because of the large fluctuations in the data. Therefore, we cannot obtain the transition temperature or the critical exponent with a systematic scaling analysis.  In Fig.~\ref{fig:macrorheo}c, we collapse the flow curves using the exponents obtained from a systematic microrheology scaling analysis in Fig.~\ref{fig:force_velo_scaling}. The collapse is not as strong as that for microrheology. Moreover, to achieve a data collapse, we adopt a higher critical temperature $T_c = 0.41$, indicating that shear rejuvenation prevents the system from reaching  equilibrium; as a result, the system falls out of equilibrium at higher temperatures.\\ 



\begin{figure}[!ht]
  \includegraphics[width=3.4in]{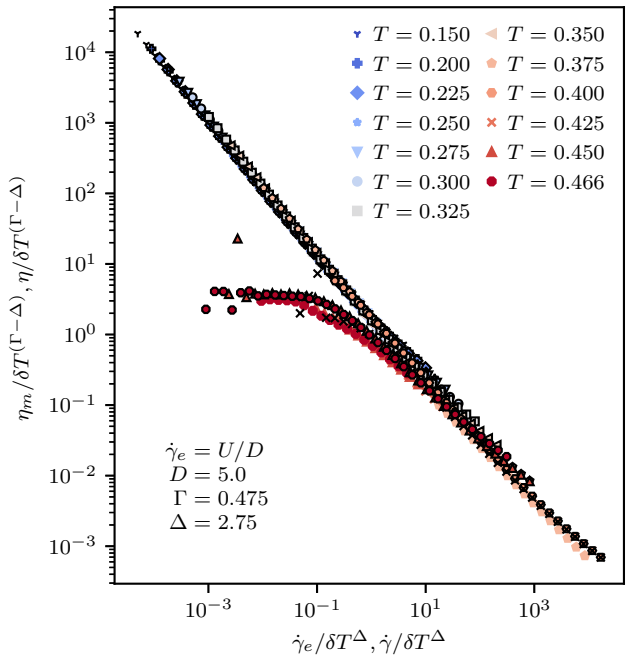}
  \caption{{\bf Superimposed $\eta$ and $\eta_m$.} 
    To bridge the gap between micro-  and macrorheology, we look at the scaling collapse of micro-viscosity ($\eta_{m}$ bold symbols) and viscosity ($\eta$) (defined in Eq.~\ref{eq:viscos}). We define an effective shear rate as $\gd_e = U/D$, where $D=2a$ is the diameter of the probe particle, for microrheology. Remarkably, the scaling collapse of viscosity  superimposes to that by micro-viscosity. This is done without any additional rescaling along X- or Y-axis.
    \label{fig:collapsed_visco}}
\end{figure}

\noindent{\bf Bridge between micro-macro rheologies.}
It is pivotal to create a link between macro- and microrheology. To this end, we seek a direct comparison of micro-viscosity ($\eta_m$) and viscosity ($\eta$), defined as
\begin{equation} 
\begin{aligned}
\eta_m &= \left(\dfrac{1}{6\pi a}\right) \dfrac{F}{U}, \\
\eta   &= \dfrac{\sxy}{\gd}
\label{eq:viscos}
\end{aligned}
\end{equation}
where $a$ is diameter of the probe particle. A scaling ansatz for micro-viscosity can be obtained from Eq.~\eqref{eq:F_U_scaling}:
\begin{equation}
\eta_m(U, T) = |T - T_{mc} | ^{\Gamma -\Delta} \mathbb{F}_\eta \left( \dfrac{U} { |T - T_{mc}| ^ \Delta} \right).
\label{eq:microvisco_scaling}
\end{equation}
A similar scaling ansatz can also be written for viscosity if $\eta_m$, $T_{mc}$, and $U$  are replaced respectively by $\eta$, $T_c$, and  $\gd$ (a formal derivation is given in the Supplementary Note~2). To compare numerical values of $\eta_m$ and $\eta$, we define an effective shear rate  $\gd_e = U/D$, where $D=2a$ is the diameter of the probe particle. This effective shear rate enables us to compare microrheology with macrorheology. Interestingly, in Fig.~\ref{fig:collapsed_visco}, the scaling collapse suggested by Eq.~\eqref{eq:microvisco_scaling} for micro-viscosity ($\eta_m / \delta T ^{\Gamma - \Delta}$ versus $\gd_e / \delta T ^{\Delta}$) superimposes on the scaling collapse for viscosity ($\eta / \delta T ^{\Gamma - \Delta}$ versus $\gd / \delta T ^{\Delta}$). This is remarkable because the micro-viscosity and viscosity superimpose without any additional rescaling along X- or Y-axis. Furthermore, this is a significant finding because (1) it provides a robust consistency of active microrheology and macrorheology after an intensive averaging over time and different realizations is done, and (2) it provides a simple recipe for direct comparison of velocity-force curves of microrheology with flow curves of macrorheology.\\



\noindent{\bf Relative width of the critical region.} In Fig~\ref{fig:macrorheo}, we qualitatively observed that width of the critical region of macrorheology is significantly smaller than that for microrheology. With $\gd_e$ and $U$ correspondence, we are now able to make a quantitative comparison between these widths. The smallest probe velocity in our simulations is $U=0.00007848$, which corresponds to an effective shear rate equal to $\gd_e=1.5\times 10^{-5}$. The critical region lies between $T = 0.300$ and $0.375$. We define the relative width of the critical region of velocity-forces curves as $\delta F = \left[ F\left(T=0.300\right) - F\left(T=0.375\right) \right] / F\left(T=0.375\right)$ at $\gd_e=1.5\times 10^{-5}$, which is equal to $\delta F = 1.13$. We now compare the relative width of the critical region ($T \in [0.375, 0.400]$) of flow curves at a shear rate approximately equal to the effective shear rate of microrheology $\gd_e$. The relative width of the critical region of flow curves can be defined as $\delta \sxy = [ \sxy(T=0.375) - \sxy(T=0.400)] / \sxy(T=0.400)$. The closet shear rate in our simulations to  $\gd_e$  is $\gd = 1.6 \times 10^{-5}$, for which we obtain $\delta \sxy = 0.14$. The ratio of $ \delta F / \delta \sxy$
reveals that microrheology gives rise to a critical region that is approximately one order of magnitude larger than that for macrorheology. This demonstrates the great potential of using microrheology for investigation of the mechanical properties of glasses and supercooled liquids.\\

\noindent{\large\bf Conclusion and discussion\par}
Here, we investigated the dynamics of a probe particle driven into a host glass forming liquid. The friction acting on the probe exhibits a second-order phase transition. The microscopic origin of friction is  one of the fundamental questions in tribology~\cite{urbakh_2004}. Our results established a non-trivial correspondence between the emergence of friction as a result of approach to glass transition. We discovered a two-step relaxation process, a characteristic of structural relaxation of a supercooled liquid, in the time auto-correlation function of the friction. Moreover, the auto-correlation function exhibits a time-velocity superposition reminiscent of the time-shear superposition principle in MCT~\cite{gotze_1992} and spin glass theory~\cite{berthier_2000}. This strengthens the relationship between dynamics of the probe particle and the structural relaxations of host medium based on existing theoretical line of argument. We further demonstrated that the auto-correlation function of friction  is a compressed exponential, shedding light on a recent controversy regarding the origin of compressed exponential relaxation in active matter~\cite{gnan_2019, angelini_2013}. Finally, we compared the results obtained by active microrheology and macrorheology, revealing that
spatiotemporaly-averaged micro-viscosity becomes equivalent to viscosity--- this is a decisive proof for the consistency of micro-macro rheologies. However, microrheology becomes superior to macrorheology because the former can provide the spatial distribution of micro-viscosity.\\


The following applications can be expected in the studies on the  microrheology for glasses or supercooled liquids. It is possible to directly observe the spatial heterogeneities of elastic modulus and viscosity in the systems. A previous study~\cite{tanguy_2002} has shown that elastic moduli  are expected to be spatially heterogeneous in glass because the response on the strain fields for weak macroscopic flow becomes heterogeneous. In the other study on supercooled liquids~\cite{furukawa_2009}, a characteristic length scale has  been observed from the wavenumber dependence of Fourier transformed  viscosity. However, there are no studies, which could obtain the spatial
distribution of elastic modulus and viscosity directly. These investigations are expected to be possible using the method of the present study.\\

The low-temperature properties of glasses have recently been the subject of intensive investigations. Originally known as the Gardner transition in spin glass theory~\cite{gardner_1985}, it has been proposed that the free energy landscape of a glass becomes fractal-like at low temperatures: a local minimum divides into many meta-basins in the free energy landscape~\cite{charbonneau_2014}. As a result of this proliferation of the meta-basins, local cages divide into several smaller cages upon cooling (or compression)~\cite{jin_2017}. Therefore, the Gardner transition is associated with an intimate change of the local morphology. As a result, a global deformation not only hinders exploration of such delicate local morphological changes but it might result in loss of the local information. This might be the reason that the existence of the Gardner transition at finite spatial dimensions has been debated~\cite{urbani_2015}. We believe that active microrheology can provide a strong experimental and computational tool for investigating the low-temperature properties of glasses, and eventually shedding light on the controversy over the existence of the Gardner transition at finite dimensions. Our work provides a firm basis for such explorations.\\

\newpage

\noindent{\large\bf Methods\par}

\noindent{\bf Simulations.} We performed large-scale $3$D simulations of a $80:20$ binary mixture  (Kob-Andersen model~\cite{kob_1995}) interacting via a smoothed LJ potential: 
\begin{widetext}
\begin{equation}
U_{ij} \left(r\right) = 
\begin{cases}
4\epsilon_{ij} \left[\left(\dfrac{\sigma_{ij}}{r}\right)^{12} -\left(\dfrac{\sigma_{ij}}{r}\right)^{6}\right] + 4\epsilon_{ij} \left[\sigma_{0ij} +  c_{2ij}\left(\dfrac{r}{\sigma_{ij}}\right)^{2}\right], \quad& r<r_{cij} \\
0, \quad& r>r_{cij} 
\end{cases}
\label{eq:LJ}
\end{equation}
\end{widetext}
where $i,j = A, B$. The parameters are represented in dimensionless LJ units in which the
length, energy, mass, and time units are $\sigma$, $\epsilon$, $m$, and
$\tau=\sqrt{\sigma^2 m /\epsilon}$, respectively. We fix the particle number density 
to $\rho = (N_A + N_B) / V = 1.2/\sigma^{3}$. The timestep of integration is $\Delta t = 0.01\tau$, and $\epsilon_{ij}$ and $\sigma_{ij}$ are set to
$\epsilon_{AA}=1.0\epsilon$, $\epsilon_{AB}=1.5\epsilon$,
$\epsilon_{BB}=0.5\epsilon$, $\sigma_{AA}=1.0\sigma$,
$\sigma_{AB}=0.88\sigma$, and $\sigma_{BB}=0.8\sigma$. Irrespective of the particle type, the mass is $m = 1$. The coefficients $c_{0ij}$ and $c_{2ij}$ are determined via  $U_{ij}(r_{cij})=0$ and $dU_{ij}/dr|_{r=r_{cij}} =0$. The cutoff, $r_{cij}$, is set to $r_{cij} = 2.5\sigma_{ij}$, and the  total number of particles is $N = 4\text{,}000$. The thermodynamic glass transition temperature for this system, the Kauzmann temperature, is $T_K = 0.3$. All simulations were performed with LAMMPS. Units are not shown explicitly throughout the text for readability unless otherwise specified\\

\noindent{\bf Preparation of initial configurations.} We prepared the initial configurations in three steps. (1) Particles are randomly inserted in the simulation box. (2) A cosine soft pair potential of the form  $U_{ij}(r)=A\left[ 1 + cos (\pi r / r_c ) \right]$ is applied to all pairs of interacting particles within the cutoff, $r_c$, regardless of the particle type. We then integrated the equations of motion of the particles using NVE integration scheme followed by NVT at $T=0.466\epsilon/k_B$. This step is crucial to minimize the overlap between particles. After minimization, the soft cosine potential  is replaced with the smoothed LJ potential given by Eq.~\eqref{eq:LJ}. (3) The system is annealed with cyclic-shearing deformation \cite{lacks_2004, das_2018} in order to obtain a stable glass that is  less prone to fluctuations. The SLLOD algorithm is used to perform non-equilibrium MD simulations with the Lees-Edwards boundary conditions \cite{evans_2008}. \\

\noindent{\bf Microrheology.} We used a singled-out probe particle to perform microrheology. The probe particle is activated by a constant velocity $U$ to mimic  constant-velocity active microrheology~\cite{zia_2018}. The probe particle interacts with the following expanded and shifted LJ pair potential:\\
\begin{widetext}
\begin{equation}
U \left(r\right) =
\begin{cases}
\infty, \quad& r \leq \Delta \\ 4\epsilon
\left[\left(\dfrac{\sigma}{r-\Delta}\right)^{12}
-\left(\dfrac{\sigma}{r-\Delta}\right)^{6}+\dfrac{1}{4}\right],
\quad& \Delta < r \leq \Delta + r_{c} \\ 0, \quad& r >
\Delta + r_{c}
\end{cases}
\end{equation}
\end{widetext}

Unless otherwise stated, $\Delta = 2.0\sigma$ and $r_c = 2^{1/6}\sigma$. Accordingly, the probe particle is 5-fold larger than the $A$-type particle. This is a typical length scale for DHs in supercooled liquids~\cite{furukawa_2009}. At $t=0$, the
probe particle is placed at the boundary, and after activation, the probe particle moves into the host at the prescribed velocity. We record the force acting on the probe particle in the direction of the velocity, $F = -{\bf F} \cdot {\bf n}$, where ${\bf n} = {\bf U} / |{\bf U}|$. Here, ${\bf F} = \sum_j {\bf f}_j $, where ${\bf f}_j$ is force exerted by a neighboring particle and the sum runs over all neighbors of the probe particle. The force is averaged over time for $500$ independent realizations. The total CPU time for the simulations is approximately  two months for a cluster of $1\text{,}656$ cores. We gathered a dataset of $20\text{TB}$ for all of the simulations.\\


\noindent{\large\bf Acknowledgment\par}
The authors thank the Korea Institute for Advanced Study for providing computing resources (KIAS Center for Advanced Computation–Linux cluster system) for this work, and especially consultations with Hoyoung Kim. This work was supported by Samsung Research Funding Center for Samsung Electronics 566 under Project Number SRFC-MA1602-02\\

\noindent{\large\bf Contributions\par}
S. H. E. R. and J. W. Y. conceived the idea. J. W. Y. performed the simulations. S. H. E. R. and J. W. Y. analyzed the data. S. H. E. R.  wrote the manuscript. All authors have discussed the results.\\

\noindent{\large\bf Competing interests\par}
The authors declare no competing financial interests.


%

\end{document}